\documentstyle[11pt,newpasp,twoside, epsf]{article}

\def\edcomment#1{\iffalse\marginpar{\raggedright\sl#1\/}\else\relax\fi}
\reversemarginpar

\begin{document}
\title{A theoretical formal model for mining transient events among databases of high energy astrophysics experiments}
\author{Francesco Lazzarotto} 
\affil{IASF-CNR via del Fosso del Cavaliere, 100 - 00133 Roma, Italy. \verb|mailto:fralaz1971@gmail.com|}
\author{Marco Feroci}
\affil{IASF-CNR via del Fosso del Cavaliere, 100 - 00133 Roma, Italy}
\author{Maria Teresa Pazienza}
\affil{DISP Tor Vergata, Via del Politecnico 1, 00133 Roma, Italy}
\begin{abstract}Data on transient events, like GRBs, are often contained in large
archives of unstructured data from space experiments, merged with
potentially large amount of background or simply undesired information.
We present a computational formal model to apply techniques of modern
computer science - such as Data Mining (DM) and Knowledge Discovering in
Databases (KDD) - to a generic, large database derived from a high energy
astrophysics experiment. This method is aimed to search, identify and
extract expected information, and maybe to discover unexpected
information.
\end{abstract}
\section{Introduction}
Giant archives of data resources are available to researchers in astronomy, and several are data not published on the web yet. (Schade et al. 2000). Two tasks are now urgent to improve knowledge extraction from our astrophysical archives:
\begin{enumerate}
\item making data available from different archives and putting them in a common and efficient way;
\item adopting the appropriate technics to extract relations in data to find new istances of known phenomena 
and to discover unknown phenomena.
\end{enumerate}
This two goals may be helped by several studies in modern computer science disciplines that come under the name of 
KDD and DM.
 
\section{Knowledge Discovery in Database (KDD) \& Data Mining (DM)}
{\bf Definition}
\begin{quote} 
\emph{The nontrivial extraction of implicit, previously unknown, and potentially useful information from data}\\ 
Frawley, W. Piatesky-Shapiro, \& G., Matheus C. J., (1992)
\end{quote}
%
%
Some outhors also name the KDD process as Data Mining (DM) but more often with DM we can refer to the analysis and 
representation of the data after to have preprocessed, cleaned and organized them.  
Our work explains how apply KDD and DM technics in order to improve astrophysical studies on GRB and, more in general, 
on high energy astrophysical transient events. We show: 

\begin{itemize}
\item the development method  of an integrated and fitting high energy astrophysical archive, which stores in a complete 
and efficient way both photon data and scientific results in a highly cross-referenced database which supports 
ad-hoc querying by users; 

\item how a formal model permits us to launch efficient explorations in order to divide the archive into event classes 
and to extract high level scientific results in an efficient and compact way. 
\end{itemize}

\begin{figure}[!htb]
\label{procKDD}
\caption{The process of Knowledge Discovery}
           \begin{center}                         
	    \plotone{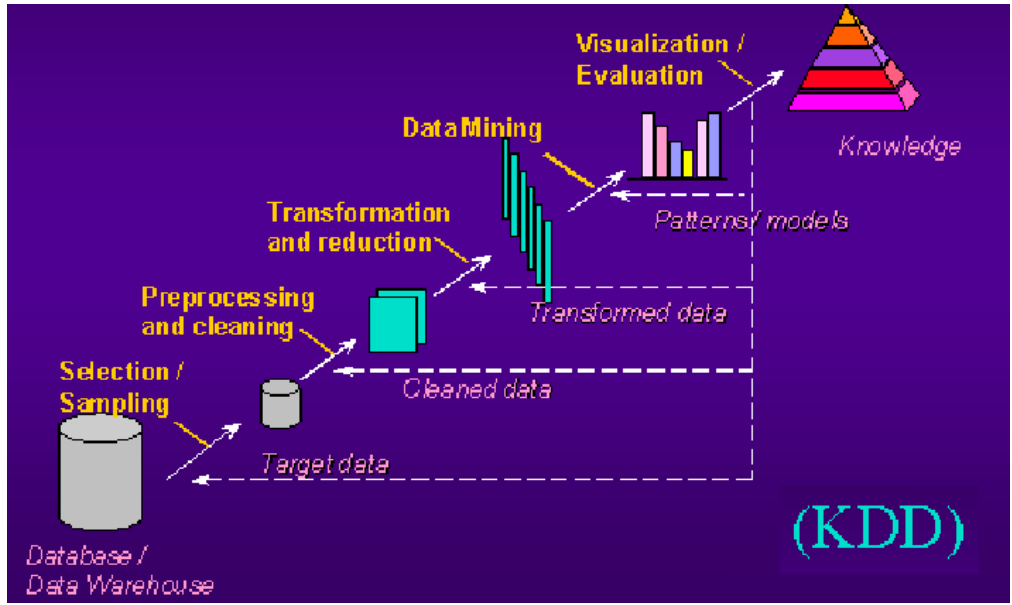}
	    \end{center}
\end{figure}

\section{The implementation of the KDD system}
\subsection{Data Warehouse}
Before performing Data Mining applications, data (better if from different databases) must be treated with preprocessing, 
cleaning and filtering operations to be stored in a standard format in an integrated database system also named 
\emph{Data Repository}.\\The whole process can assume the name of \emph{Data Warehousing}  	
\begin{figure}[!htb]
\caption[Data Warehouse]{Data Warehousing step}
\begin{center}              
  \plotone{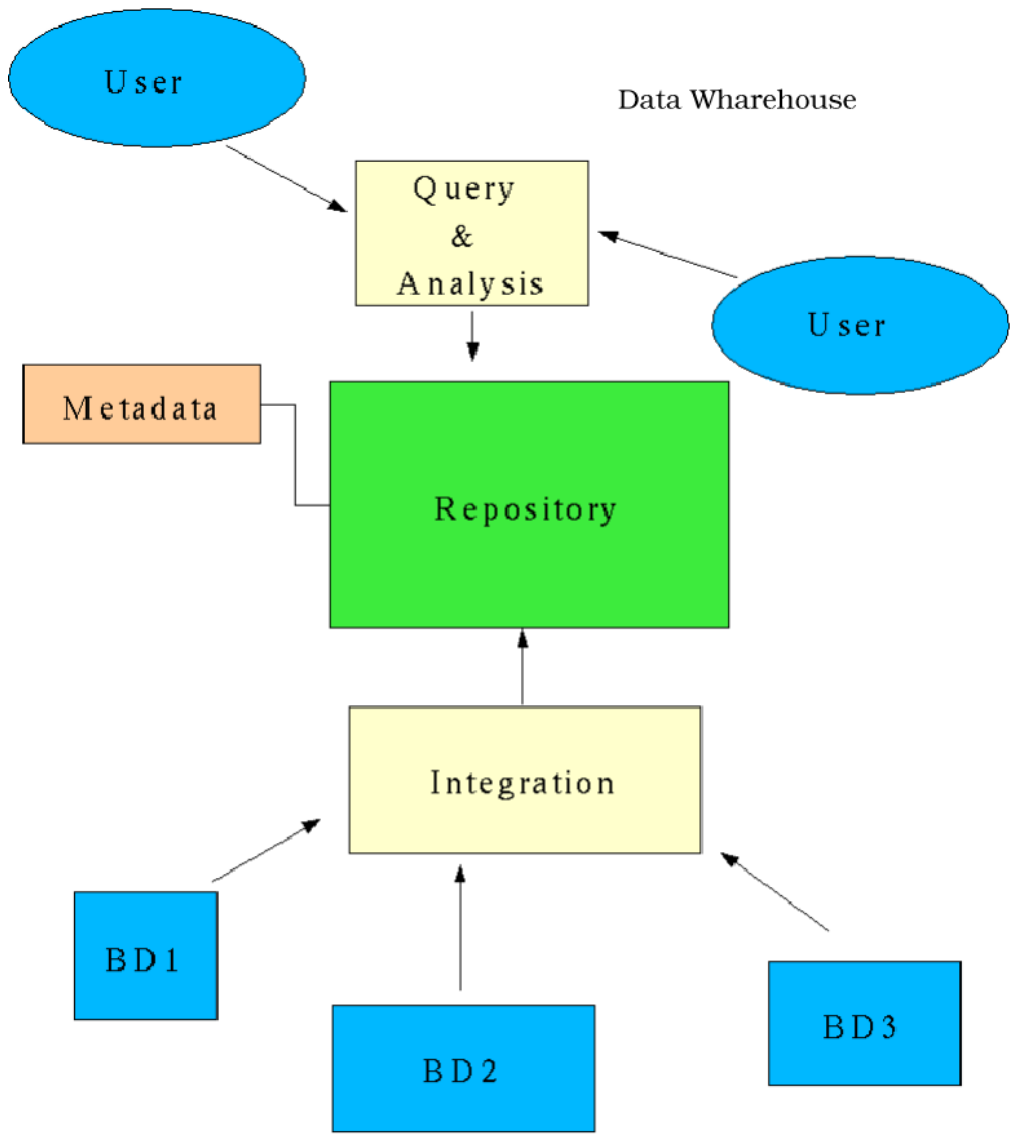}
\end{center}                 
\label{DWproc}
\end{figure}   
            
                  
\subsection{The data organization}
To perform an efficient information retrieving amonog GRB data sets, cannot be disregarded the use of instruments 
that permit to submit simple queries, based on a formal or natural language, the minimum is to implement a 
Data Base Management System(DBMS). SQL (Simple Query Language) is also suitable, but for very large databases 
would be better to use the OLAP (On Line Analytical Processing) approach based on integrated 
hierarchic and multidimensional representation of data.
        
\section{Mining GRB among astrophysics databases}
Preliminary studies (Feroci et al. 1999; Lazzarotto 2001) highlighted the segmentation of SAX GRBM detector 
on-ground database, in different classes of events. 

\begin{figure}[!htb]
\vspace{2.5cm}
\caption[Signals Classes]{Classes of signals}
\begin{center}              
		\plotone{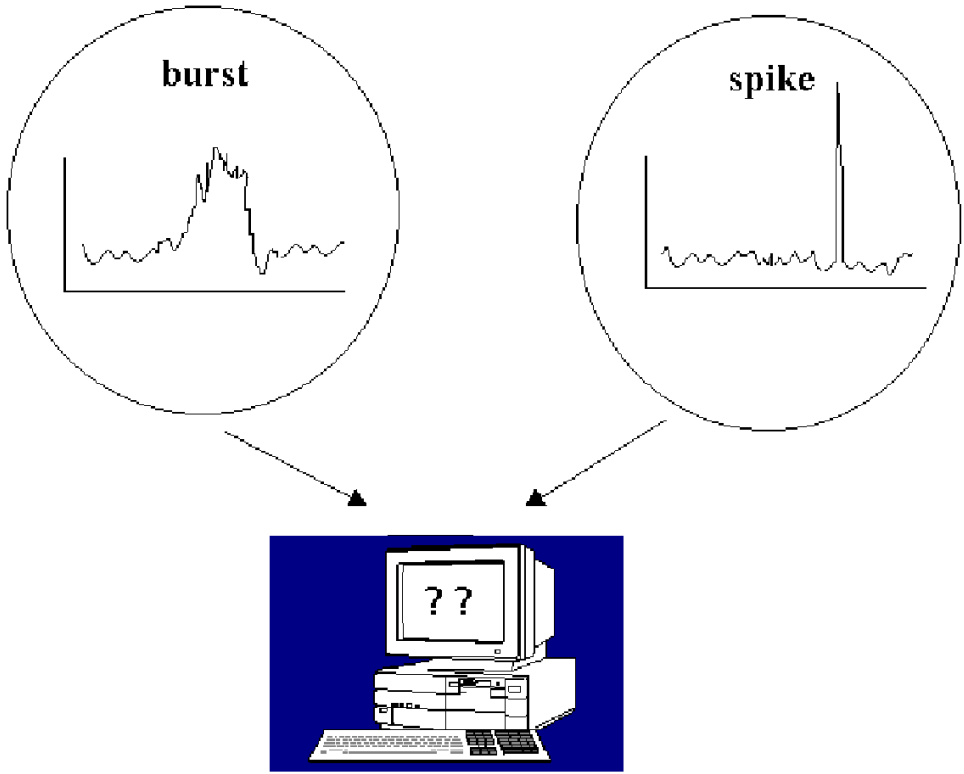}
\end{center}             
\label{Signals}
\end{figure} 
       
\subsection{Preprocessing and cleaning data}
Noisy and corrupted data are a large part of the data, during extensive analysis on IASF/CNR GRBM archive since 1999, 
were been detected and corrected many causes of errors in the data such as:
\begin{itemize}
\item space acquisition and transmission errors;
\item on-ground preprocessing errors;
\item local software errors.
\end{itemize}

\subsection{Fuzzy logic, Clustering and Pattern recognition}
In the early analysis (Lazzarotto 2001) we used a self-implemented object oriented database, that realized 
only queries and statistical global operations we thought were significative, now we intend to implement 
a standard KDD system based on a DW in order to apply more efficient tecnichs of DM. In the past works, 
we used \emph{fuzzy logic} to perform pattern recognition and to discriminate different kinds of presumibly known signals. 
 
\section{Steps for the model}
\begin{figure}[!htb]
\caption[GRBM archive]{Composition of the GRBM instrument archive}
	\begin{center}              
 	      \plotone{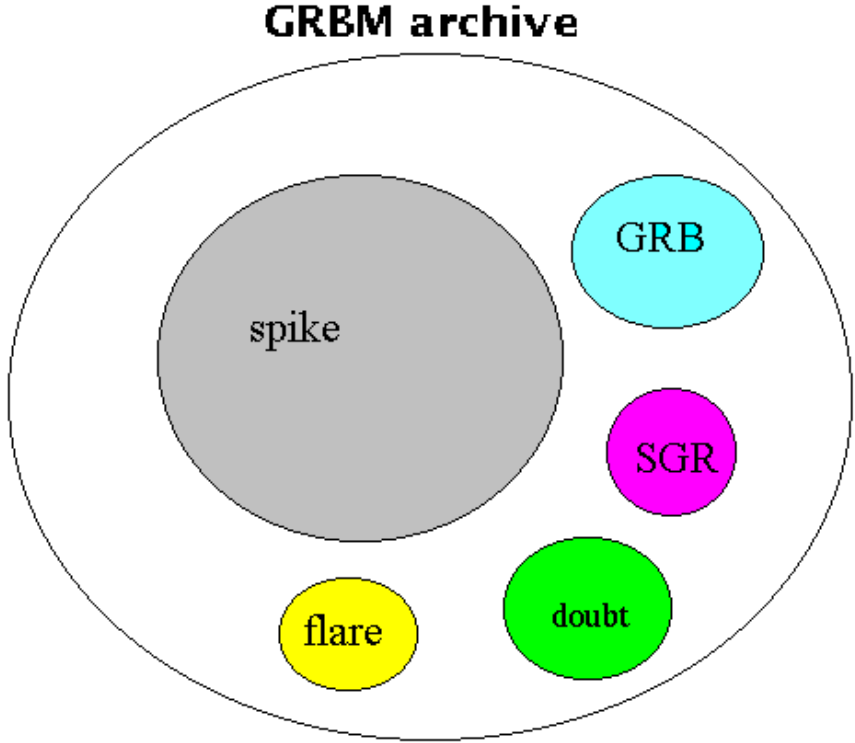} 
       \end{center}
\label{GRBMarchive}
\end{figure}   

The following steps show how to improve the ability of filtering events.
\begin{enumerate}
\item Definition of quantitative data and attributes characterizing a transient event (event list, duration in seconds, 
location, spectral measures, flux, \ldots) 
\item Definition of categorical attributes characterizing a transient event (hardness, class of duration, 
\#instruments that have detected it, flux level, \ldots)
\item application of 3 basical DM tecnichs:
\begin{itemize}
\item discovering of association rules;
\item cluster analysis;
\item temporal series analysis. 
\end{itemize}
\end{enumerate}
In Feroci et al. (1999); Lazzarotto (2001), we went deep basically on temporal series to make a partition 
of the data set into known classes. A complete KDD system permits to improve past work and to apply 
the predictive technics we want to show. 
\subsection{Associations among event attributes}
\begin{itemize}
\item We define an event as a \emph{cube} of observations ($O_i$) for that event, temporally divided in 
\emph{Peaks} ($P_j$) with some \emph{Attributes} ($A_k$) 
\end{itemize} 
Let be an event $E=\{A_1, \ldots, A_k\}$  a set of categorical attribute instances. Let be $D$ a set of events.

{\bf Association rule} $R$: $(A_m=x) \longrightarrow (A_n=y)$ with $(A_m, A_n \in E)$ 
\begin{itemize}
\item Support = $Supp_{(A_m=x) \rightarrow (A_n=y)} = \frac{|E_i \in D : (A_m=x) \land (A_n=y)|}{|D|} $ 
\item Confidence = $Conf_{(A_m=x) \rightarrow (A_n=y)} = \frac{|E_i \in D :(A_m=x) \land (A_n=y)|}{|E_i:(A_m=x)|}$ 
\end{itemize}
{\bf Problem }: find all the rules $R$ in a transient events dataset D : \\$(Supp_R \ge minSupp) \land (Conf_R \ge minConf )$

\begin{figure}[!htb]
\caption[Event cube]{formal and visual model for a transient event}
	\begin{center}              
\plotone{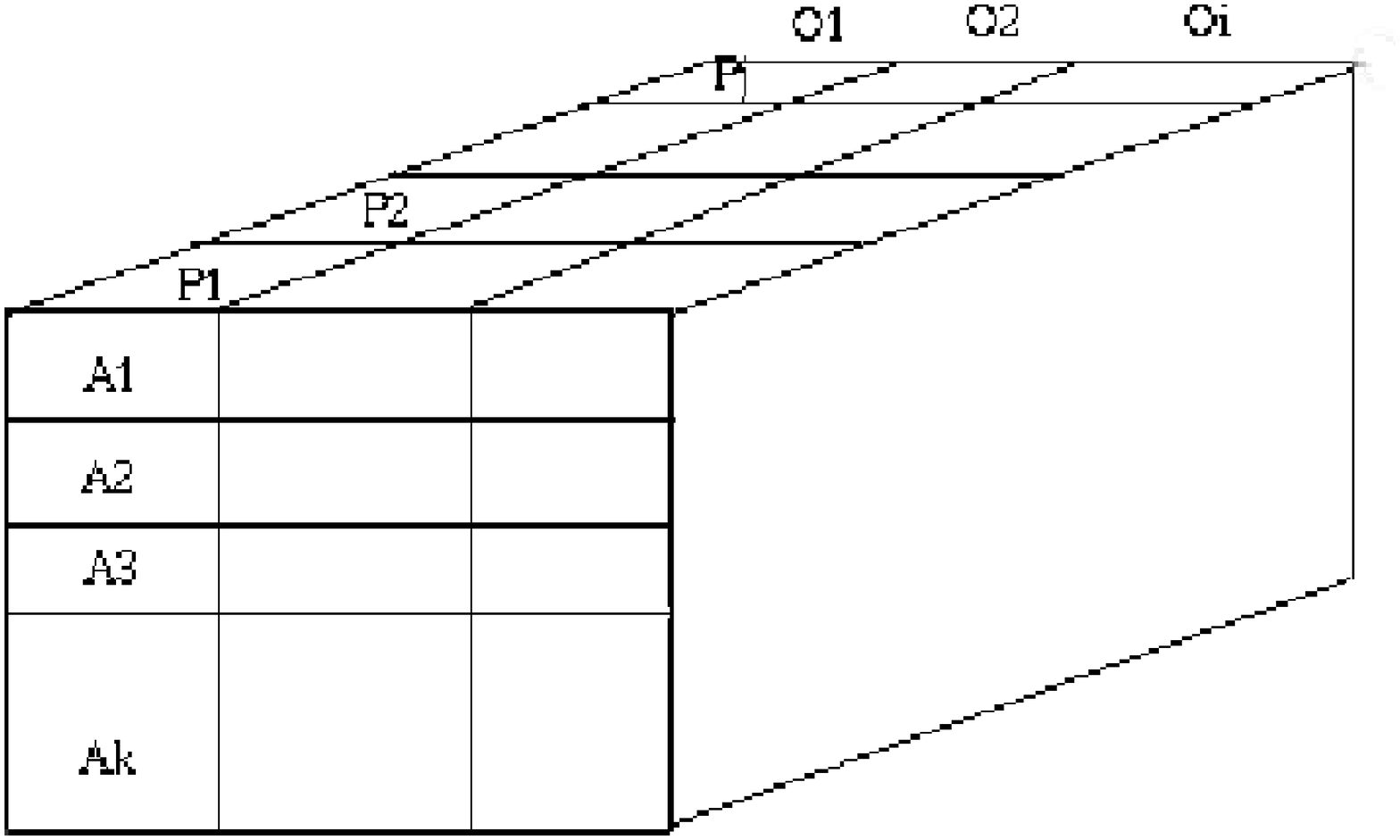}
	\end{center}  
\label{EventCube}
\end{figure}   

\subsection{Density based clustering of events}

\begin{figure}[!htb]
\caption[Clustering]{Clustering algorithms to discovery burst classes}
\begin{center}              
\plotone{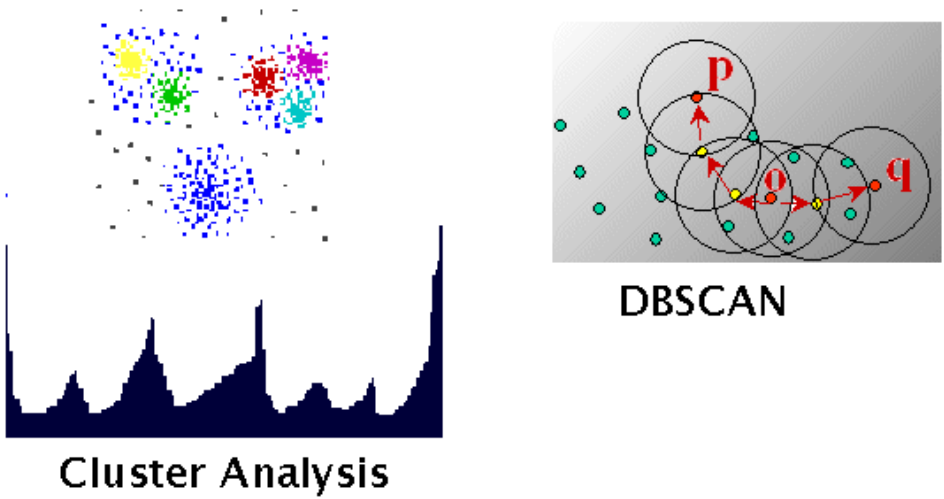}
\end{center}
\label{DBSCAN}
\end{figure} 

Another approach to knowledge extraction among a large and noisy transient events database, is to adopt \emph{clustering algorithms}. 
We chose to apply a density-based clustering algorithm. The concept is to think that clusters of events are dense regions 
of a multidimensional space distiguished from sparse regions that represent the \emph{noise}. 
These algorithms need an event defined as a set of attributes $E=\{A_1, \ldots, A_k\}$ that respects the axioms 
of a \emph{metric space}. The basical idea: to decide if an event is in a cluster, we have to find that the density 
of events in the \emph{neighborhood} of that event, must exceed a certain threshold. 
A fundamental work in this field is Density Based Spatial Clustering of Application with Noise 
(DBSCAN by Ester, Kriegel, Sander \& Xu, 1996). 
\section{Conclusions}
\begin{itemize}
\item We defined a basical criterium to select a transient event from a temporal series 
(i.e. T90 as we made in (Feroci et al. 1999; Lazzarotto 2001)
\item Applying correctly defined attributes of events: duration, position, hardness, shape (rising/falling front, FWHM), 
we have now the correct methodological system and theorical instruments to launch our next analysis. 
\item We have to spend a startup time to implement the global system, engeneering actual tools and algorithms, 
and to train the system to known results.
\item Then we can launch the system on large datasets, in order to find classifications among events and relations among 
events attributes, without supposing criteria that always result not general and imprecise, when we change or enlarge 
our datasets of hypotetical GRB events. 
\item We can quickly change the point of interest, change kind of data with a minimum work, have facilities to represent 
results. 
\end{itemize}
And this is very important in a mysterious problem like GRBs, where lots of instrumental data are not globally analyzed yet.
\section{Acronyms}
\begin{description}
\item[DB] DataBase.
\item[DBMS] Data Base Management System.
\item[DM]  Data Mining.
\item[DW] Data Warehousing.
\item[FWHM] Full Width Half Maximum
\item[GRB] Gamma Ray Burst.
\item[GRBM] Gamma Ray Burst Monitor.
\item[KDD] Knowledge Discovery in Database.
\item[OLAP] OnLine Analytical Processing.
\item[SAX] Satellite for Astronomy in X rays.
\item[SGR] Soft Gamma ray Repeater.
\item[SQL] Simple Query Language.
\end{description}

\end{document}